\documentclass
[twocolumn,showpacs,preprintnumbers,amsmath,amssymb]
{revtex4}
\usepackage{graphicx}
\usepackage{dcolumn}
\usepackage{bm}
\usepackage{here}
\usepackage{color}
\usepackage{braket}
\usepackage{ascmac}

\begin{document}

\newcommand{\vc}[1]{\mbox{\boldmath $#1$}}
\newcommand{\fracd}[2]{\frac{\displaystyle #1}{\displaystyle #2}}
\newcommand{\red}[1]{\textcolor{red}{#1}}
\newcommand{\blue}[1]{\textcolor{blue}{#1}}
\newcommand{\green}[1]{\textcolor{green}{#1}}



\def\ni{\noindent}
\def\nn{\nonumber}
\def\bH{\begin{Huge}}
\def\eH{\end{Huge}}
\def\bL{\begin{Large}}
\def\eL{\end{Large}}
\def\bl{\begin{large}}
\def\el{\end{large}}
\def\beq{\begin{eqnarray}}
\def\eeq{\end{eqnarray}}
\def\beqnn{\begin{eqnarray*}}
\def\eeqnn{\end{eqnarray*}}

\def\bit{\begin{itemize}}
\def\eit{\end{itemize}}
\def\bsc{\begin{screen}}
\def\esc{\end{screen}}

\def\eps{\epsilon}
\def\th{\theta}
\def\del{\delta}
\def\omg{\omega}

\def\e{{\rm e}}
\def\exp{{\rm exp}}
\def\arg{{\rm arg}}
\def\Im{{\rm Im}}
\def\Re{{\rm Re}}

\def\sup{\supset}
\def\sub{\subset}
\def\a{\cap}
\def\u{\cup}
\def\bks{\backslash}

\def\ovl{\overline}
\def\unl{\underline}

\def\rar{\rightarrow}
\def\Rar{\Rightarrow}
\def\lar{\leftarrow}
\def\Lar{\Leftarrow}
\def\bar{\leftrightarrow}
\def\Bar{\Leftrightarrow}

\def\pr{\partial}

\def\>{\rangle} 
\def\<{\langle} 
\def\RR {\rangle\!\rangle} 
\def\LL {\langle\!\langle} 
\def\const{{\rm const.}}

\def\e{{\rm e}}

\def\Bstar{\bL $\star$ \eL}

\def\etath{\eta_{th}}
\def\irrev{{\mathcal R}}
\def\e{{\rm e}}
\def\noise{n}
\def\hatp{\hat{p}}
\def\hatq{\hat{q}}
\def\hatU{\hat{U}}

\def\hatA{\hat{A}}
\def\hatB{\hat{B}}
\def\hatC{\hat{C}}
\def\hatJ{\hat{J}}
\def\hatI{\hat{I}}
\def\hatP{\hat{P}}
\def\hatQ{\hat{Q}}
\def\hatU{\hat{U}}
\def\hatW{\hat{W}}
\def\hatX{\hat{X}}
\def\hatY{\hat{Y}}
\def\hatV{\hat{V}}
\def\hatt{\hat{t}}
\def\hatw{\hat{w}}

\def\hatp{\hat{p}}
\def\hatq{\hat{q}}
\def\hatU{\hat{U}}
\def\hatn{\hat{n}}
\def\hatb{\hat{b}}
\def\hata{\hat{a}}
\def\hath{\hat{h}}

\def\hatphi{\hat{\phi}}
\def\hattheta{\hat{\theta}}

\def\hJ{\hat{J}}
\def\hI{\hat{I}}
\def\hphi{\hat{\phi}}
\def\hp{\hat{p}}
\def\hq{\hat{q}}
\def\hU{\hat{U}}
\def\hX{\hat{X}}
\def\hR{\hat{R}}

\def\hH{\hat{H}}
\def\hh{\hat{h}}
\def\hW{\hat{W}}

\def\Z{\mathbb{Z}}
\def\ovl{\overline}

\def\iset{\mathcal{I}}
\def\fset{\mathcal{F}}
\def\pr{\partial}
\def\traj{\ell}
\def\eps{\epsilon}

\def\U{U_{\rm cls}}
\def\P{P_{{\rm cls},\eta}}
\def\traj{\ell}
\def\cc{\cdot}

\def\DZ{D^{(0)}}
\def\Dcls{D_{\rm cls}}

\newcommand{\relmiddle}[1]{\mathrel{}\middle#1\mathrel{}}

\title{Quantum Diffusion Induced by Small Quantum Chaos}
\author{Hiroaki S. Yamada}
\affiliation{Yamada Physics Research Laboratory,
Aoyama 5-7-14-205, Niigata 950-2002, Japan}
\author{Kensuke S. Ikeda}
\affiliation{College of Science and Engineering, Ritsumeikan University, 
Noji-higashi 1-1-1, Kusatsu 525-8577, Japan}

\date{\today}
\begin{abstract}
%
It is demonstrated that quantum systems classically exhibiting strong and homogeneous
chaos in a bounded region of the phase space can induce a global quantum 
diffusion. As an ideal model system, a small quantum chaos with finite Hilbert 
space dimension $N$ weakly coupled with $M$ additional degrees of freedom
which is approximated by linear systems is proposed.
By twinning the system the diffusion process in the additional modes can
be numerically investigated without taking the unbounded diffusion space 
into account explicitly.
%
Even though $N$ is not very large, diffusion occurs in the additional modes as 
the coupling strength increases if $M \geq 3$. 
If $N$ is large enough, a definite quantum transition to diffusion takes place
through a critical sub-diffusion characterized by an anomalous diffusion 
exponent. 
\end{abstract}



\maketitle


{\it Introduction.-}
By introducing any perturbation to completely integrable systems, chaotic 
region is formed close to the nonlinear resonance which exists almost 
everywhere in the phase space. However, chaotic components are prevented 
to globalize by the KAM tori and are localized \cite{lichtenberg92,chirikov79}.
However, if such a small 
localized chaos interacts with some additional degrees of freedom, 
it can drive them and can change their energies on a large scale.

A typical example is the mechanism proposed by Arnold \cite{arnold64}.
He showed that the entanglement between the stable and unstable manifolds 
of the unstable fixed point of a resonance, which causes the so called 
stochastic layer chaos, simultaneously leads to the intersection of stable 
and unstable manifolds with different energies of 
the additional degrees of freedom, thereby forming a global path to 
change their energy. Such kind of global instability is called the Arnold 
diffusion and were treated analytically and numerically 
\cite{nekhoroshev77,chirikov79,lieberman80,holmes82,kaneko85}.
The global motion induced by a localized small chaos such as 
the stochastic layer is an initiation leading to the intrinsically 
global ergodic motions 
\cite{chirikov79,cincotta14,santhanam22,maurya22}.

Investigations of quantum Arnold diffusion for various systems 
elucidated that quantum motion mimics the classical delocalization.
 \cite{wolynes97,demikhovskii02a,demikhovskii02b,boretz16}.
However, the diffusion rate is much smaller than the classical one
and very long-time behavior of the quantum diffusion is not known.
It is expected to be suppressed by the quantum localization effect 
\cite{demikhovskii02a,demikhovskii02b,demikhovskii06}. 

So far the quantum chaotic diffusion has been investigated extensively 
for ``large'' quantum chaos system defined in unbounded phase space with
infinite Hilbert-space dimension \cite{casati79,fishman82,casati89,chotorlishvili19}.
 In such systems chaotic degrees of freedom 
itself may actively exhibit diffusion  in the classical limit, but
the diffusion is inhibited in its quantum counterparts due to the quantum 
localization effect. Such quantum localization is, however,
destroyed,  and classical diffusion is recovered \cite{adachi88b,gadway13}
through Anderson-like transition
if the number of degrees of freedom increases \cite{casati89,lopez12,lopez13,yamada20,chotorlishvili19}.
On the other hand, the nature of quantum diffusion {\it passively induced by 
small quantum chaos system} have not been known, although it is a fundamental 
problem closely related to the quantum global instability such as the Arnold diffusion.
In the present paper, we propose a simple model and method, 
with which we can examine whether or not a small quantum chaos 
can induce global diffusive motion along the modes contacting with 
it, and show that a global transportation is realized under appropriate 
conditions.
“Small chaos” means chaotic systems confined to a finite region 
of the phase space by geometrical or dynamical conditions in the 
classical limit. As a first step, we examine small 
but sufficiently unstable chaos.  Small and weakly unstable 
chaos such as stochastic layer will be investigated in forthcoming papers.

{\it Model.-}
As the first class of example, we consider strong and uniform chaotic systems
called as C-system or K-system
bounded finitely by periodic boundaries \cite{lichtenberg92},
which are coupled with several number of unbounded additional 
degrees of freedom. 
The former is referred to as the main system 
and latter as the additional modes, respectively.
The additional modes are supposed to be integrable if isolated, as is the 
example considered by Arnold.
They are coupled weakly with the former at
 a small coupling strength characterized 
by the parameter $\eta$. 

We suppose the integrable additional modes of number $M$ is initially 
located at the action eigenstate $|I_{0k}\>~(1\leq k\leq M)$
where $I_{0k}=\hbar\times$integer. We approximate 
the Hamiltonian of the additional modes by linearizing around $I_k\sim I_{0k}$, 
and reset $\hI_k-I_{0k}$ by $\hJ_k$. Then  Hamiltonian of  the entire system is represented by 
\def\hHeff{\hat{\cal H}}
\def\hUeff{\hat{\cal U}}
\def\hHeffT{\hat{\cal H}^T}
\def\hUeffT{\hat{\cal U}^T}
\def\hU{\hat{U}}

\beq
\label{eq:H}
\hH(\hp,\hq, \hJ, \hphi, t)&=&\hHeff(\hp,\hq,\hphi,t)+\hh(\vc{\hJ}) \\
\nn \hHeff(\hp,\hq,\hphi,t)&=&\hp^2/2+V(\hq)\Delta(t)+\eta v(\hq)w(\vc{\hphi})\Delta(t),
\eeq
where $\Delta(t):=\sum_{n\in\Z}\delta(t-n)$ is the periodic delta-functional kicks.
$\hh(\vc{\hJ})=\vc{\omega\hJ}$, where $\vc{\hJ}=(\hJ_1,..,\hJ_M)$, 
and $\vc{\omega}=(\omega_1,..,\omega_M)$ are the 
the linear frequencies at $I_k=I_{0k}(k=1,...,M)$.
They are supposed to be mutually incommensurate. 
 $\vc{\hphi}=(\hphi_1,...,\hphi_M)$ are the angle operators conjugate to action operators 
$\hJ_k$ as $\hJ_k=-i\hbar \pr/\pr \phi_k$ in the c-number representation of 
$\hphi_k~(0\le\phi_k\le 2\pi)$. 
We take $w(\vc{\hphi}):=\sum_{k=1}^Mw(\hphi_k)$, 
where  $w(\phi_k)$ is a $2\pi$ periodic function of angle variable $\phi_k$ 
with mean value 0.

Here, we take the main system represented by as a 
kicked rotor driven by the periodic kick $\Delta(t)$ of period 1
applied to the potential $V(\hq)$, where $\hq=\sum_qq|q\>\<q|$ and
$\hp=\sum_pp|p\>\<p|$ are the position and momentum operators
with eigenvalues $q$ and $p$, respectively.


The main system is confined in the bounded phase 
space $-\pi\leq p\leq \pi,~-\pi \leq q\leq \pi$, and the periodic boundary 
conditions are imposed on $p$ and $q$. Then they are quantized as 
$q=\ell \hbar$ and $p=\ell'\hbar$,  where $\ell, \ell'$ are the integers 
satisfying  $-N/2\leq \ell,\ell' \leq N/2$ with $N$ being the Hilbert-space 
dimension of the main system related to the Planck constant as $\hbar=2\pi/N$.

Next, we take the Arnold cat map $V(\hq)=K\hq^2/2$ 
or the standard map $V(\hq)=K\cos \hq$ 
defined in the above bounded phase space as the main system $H_0(\hp,\hq,t)$.
Taking $K(\in\Z)$ as $K>4$ or $K<0$(cat map) or $|K|\gg 1$(standard map) the 
main system can be made C-system and approximately K-system, 
respectively, which are (almost) uniformly chaotic with a flat invariant measure 
in the classical limit.
The interaction terms $v(q)$ and $w(\phi)$ are period $2\pi$-functions of $q$ and $\phi$ ,
respectively, with $0$ mean for the uniform invariant measure.
We choose $v(q)=\cos(q)$ and $w(\phi_k)=\cos\phi_k$ of the interaction term
 in this paper.
 The similar model with linear oscillators have been used 
 by several authors while studying the chaotic dynamics of the rotors 
 \cite{shepelyansky83} and 
 Anderson transition of the atomic matter waves \cite{chabe08,lemarie09}.


A great merit of using the linear oscillators as the additional mode is that the
unitary evolution operator $\hU(t)={\cal T}\exp\{-i\int_0^t \hH(\hp,\hq,\hJ,\hphi,s)ds/\hbar\}$
can be factorized into the action-dependent part and angle-dependent part as
\beq
\label{eq:UandUeff}
\nn  \hU(t)&=&e^{-i\hh(\vc{\hat{J}})t/\hbar}\hUeff(t,\vc{\hphi}), \\
 \hUeff(t,\vc{\hphi})&:=& {\cal T}e^{-\frac{i}{\hbar} \int_0^t \hHeff(\hp,\hq,\vc{\hphi}+\vc{\omega}s)ds},
\eeq
where ${\cal T}$ means the time-ordering operator.
If the operator $\hX$ does not contain the angle operators, 
the time evolved operator $\hX(t)=\hU^{\dagger}\hX \hU$
is dominated by $\hUeff(t)$ as $\hX(t)=\hUeff(t)\hX\hUeff(t)$. 
The action $\hJ(t)$ changes only at the $t=n$($n \in \Z$)-th kick. The Heisenberg equation of motion 
$d\hJ/dt=i[\hHeff,\hJ]/\hbar$ is integrated at each kick to leads to
\beq                 
\label{eq:DJ}
 \hJ_k(t)-\hJ_k(0)= -\sum^{[t]}_{n=1} \eta v(\hq(n)) w'(\hphi_k+\omega_k n), 
\eeq
where $[~~]$ is the Gauss symbol and $w'(\phi_k):=dw(\phi_k)/d\phi_k$. 
Our interest is whether or not the chaotic motion of the main system can induce
a global transport in the action space starting from $|\vc{J}=0\>$. 
The physical quantity directly measuring the transported distance is 
the mean square displacement (MSD) of the action:
$\Delta J_k(t)^2 :=\<\Psi_0|(\hJ_k(t)-\hJ_k(0))^2|\Psi_0\>$ $(k=1,2...,M)$, 
where $|\Psi_0\>=|\psi_0\>|\vc{J}=0\>$
and $|\psi_0\>$ is the initial state of the main system. 
Classically, Eq.(\ref{eq:DJ}) describes a typical situation in 
which chaos induces diffusion in the additional modes:if the main system is fully chaotic, 
the 'force' $v(q(n))w'(\phi_k+\omega_k n)$ is completely random with $0$ mean, 
and the classical variable $\Delta J_k(t)$ exhibits a Brownian motion
To compute the MSD or higher-order moment, we need not explicitly taking the 
infinite Hilbert dimension of the $J$-space into account as shown below.\\

\begin{figure}[htbp]
\begin{center}
\vspace{-5mm}
\includegraphics[height=3.6cm]{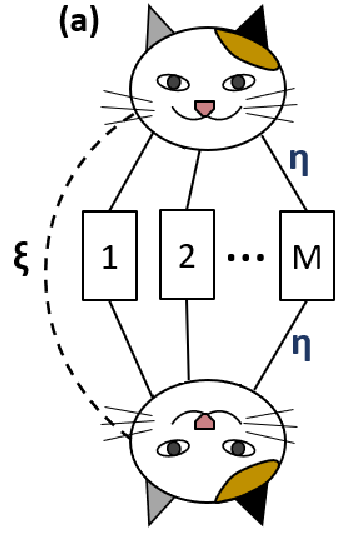}
\hspace{1mm}
\vspace{-1mm}
\includegraphics[height=4.0cm]{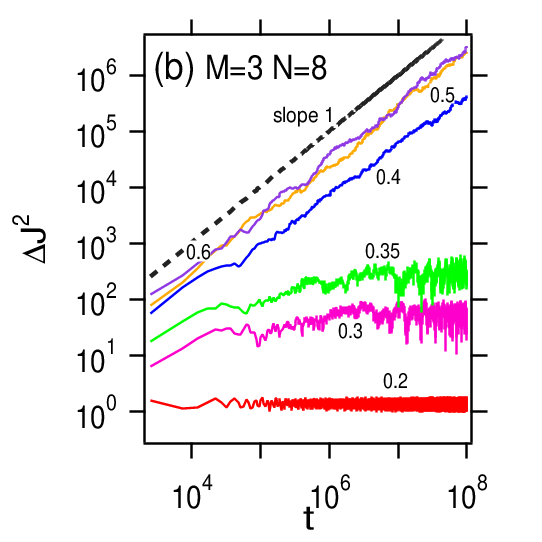}
\caption{(Color online) \label{fig:N=8}
(a) An Illustration of the twinned system with the coupling strength $\eta$ and $\xi$.
(b)The double-logarithmic plots of $\Delta J^2(t)$ as a function of time
for the perturbed cat map  for $K=-1$ of $N=8$
with trichromatic perturbation ($M=3$).
The result for various $\eta$ in the range $\eta \in[0.2, 0.6]$ is shown.
}
\end{center}
\end{figure}

{\it Method.-}
We twin the two identical parts represented by the Hamiltonian
$\hHeff(\hp,\hq,\vc{\hphi},t)$ of Eq.(\ref{eq:H}) and its paired one $\hHeff(\hp',\hq',\vc{\hphi},t)$.
Returning $t$ to the continuous time representation, the Hamiltonian $\hH_{T}$ of the 
twinned system is 
\beq
\label{eq:H2}
 \nn && \hH^T_\xi= \hHeffT_\xi(\hp,\hq,\vc{\hphi},t)+h(\vc{\hJ})\\
     && \hHeffT_\xi:=\hHeff(\hp,\hq,\vc{\hphi},t)+\hHeff(\hp',\hq',\vc{\hphi},t)+ \xi\hW(\hphi_k,t).
\eeq
The second part is spanned by the coordinate basis $|q\>'$ or the momentum basis
$|p\>'$, and $\hq':=\sum_q q|q\>'\<q|'$  and $\hp':=\sum_p p|p\>'\<p|'$ are
its coordinate and momentum operators, respectively.
$\hW$ is the interaction between the twinned parts given by
\beq
 \hW(\hphi_k,t):= w'(\hphi_k)\sum_qv(q)(|q\>\<q|'+|q\>'\<q|)\Delta(t-0), 
\eeq
The twin system is illustrated in Fig.\ref{fig:N=8}(a). 
The interaction between the twins takes place just at $t=n+0(n\in\Z)$ 
after the periodic kick.
We define here the transition operators $\hR^+=\sum_q|q\>'\<q|$ and $\hR^-=(\hR^+)^\dagger$.
Since our the system is formed by twins and the non-interacting Hamiltonian $\hH^T_{\xi=0}$ 
commutes with $\hR^{\pm}$, the time-evolved operators $\hR^{\pm}$ change
only in the moments of the interaction at $t=n+0$. Let $\hU^T_\xi(t)$ and $\hUeffT_\xi(t)$ be
the time evolution operator of the Hamiltonian 
$\hH^T_\xi$ and $\hHeffT_\xi(\hp,\hq,\vc{\hphi}+\vc{\omega}t)$, 
respectively. Then the  relation similar to Eq.(\ref{eq:UandUeff}) holds 
and $\hR^{\pm}(t)$ is dominated by $\hHeffT_\xi(\hp,\hq,\vc{\hphi}+\vc{\omega}t)$. 
As a result, we obtain
\beq
\label{eq:R}
& &  (\hR^+(t)-\hR^+(0))/\xi \nn \\
&=&\frac{i}{\hbar}\sum_{s=1}^{[t]} [v(\hq(s))-v(\hq'(s))]w\red{'}(\hphi_k+\omega_k s).
\eeq
We suppose initially the second chaotic system is not populated, and only the first main system
and the additional modes are started from the same initial condition as the original system 
given by Eq.(\ref{eq:H}). Here the coupling strength $\xi$ 
of the twinned system is chosen at the smallest level within 
the numerical precision allows. 
The RHS of Eq.(\ref{eq:R}) is computed in the lowest order with respect to $\xi$.
Then the second term of RHS, which contains only the population operator 
of the second system, can be neglected, and the LHS of Eq.(\ref{eq:R}) is identified 
with $\hJ_k(t)-\hJ_k(0)$.
The MSD is thus related to the excitation number $\hR^+\hR^-$ as
\beq
\label{eq:MSD}
& & \Delta J_k(t)^2:= \<\Psi_0|(\hJ_k(t)-\hJ_k(0))^2|\Psi_0\> \nn \\
&=&\<\Psi_0|(\hR^+(t+0)-\hR^+(0))(\hR^-(t+0)-\hR^-(0))|\Psi_0\>/\xi^2 \nn \\
&\simeq&
\frac{\hbar^2 \eta^2}{(2\pi)^M\xi^2}\int d\vc{\phi}\sum_{q}|\<q|'\hUeffT_\xi(t,\vc{\phi})|\psi_0\>|^2,
\eeq
where $\int d\vc{\phi}=\int_0^{2\pi}...\int_0^{2\pi}\Pi_k^Md\phi_k$.
As mentioned above, 
$\hUeffT_\xi(t, \vc{\phi})
={\cal T}\exp\{-i\int_0^t \hHeffT_\xi(\hp,\hq,\vc{\phi}+\vc{\omega}s,s)ds/\hbar\}$, 
which can be expressed as the product of step-by-step evolution operator.
More detailed derivation and remarks about Eqs.(\ref{eq:R}) and (\ref{eq:MSD})
can be found in the Appendix \ref{app:twin-method}.
In the following we omit $k$ from $\hJ_k$ and $J_k$ if not necessary.

Thus the MSD is copied to the total excitation number of the second main system.
The higher-order moment can also be evaluated in the same way.
The integration over the phase variables $\vc{\phi}$ means to take quantum mechanical
average with respect to the initial action state 
$|\vc{J=0}\>=\int_0^{2\pi}...\int_0^{2\pi}d\vc{\phi}|\vc{\phi}\>/(2\pi)^{M/2}$
which is very efficiently carried out by replacing the integral by the average
over quasi random numbers of the integer$(\nu)$-multiplied irrational number 
$\phi_k=\nu\chi_k (0\leq \nu \leq \nu_{max})$, where $\chi_k$ are irrational numbers.
   
We have only to execute the numerical wavepacket evolution with 
$\hUeffT_{\xi}(t,\vc{\phi})$ using $2N$-dimensional basis for fixed c-number 
$\phi_k=\nu\chi_k$ starting from $|\psi_0\>$, and compute 
the integrand of Eq.(\ref{eq:MSD}) for a fixed $\phi_k=\nu\chi_k$,
and next take the average over the $\nu_{max}$ data. 
Finally the average over the results for randomly chosen initial state
$|\psi_0\>$ is taken. 

We compared the result of the twinning method with the result of the direct 
wavepacket propagation in the full Hilbert space spanned by the truncated 
set of action basis and the $N$-dimensional basis of main system.
The results agree well, 
which are demonstrated in the Appendix \ref{app:twin-method}. 

{\it Result.-}
We first take the chaotic cat map as the main system, 
which induces ideal Diffusion process
according to Eq.(\ref{eq:DJ}) in the classical model. 
However in the quantum version 
follows the classical chaotic diffusion at least in a certain period of time evolution. 
Indeed, in the case of $M=1$, the diffusion is suppressed and the MSD reaches to an upper
bound $\ell^2:=\Delta J^2(t=\infty)$
(for brevity the index $k$ is omitted).
We call $\ell$ the quantum localization length. 
Numerically $\ell \propto N^2D_{cl}$, where $D_{cl}$  
is classical diffusion constant proportional to $\eta^2$, 
which is predicted by
a simple theoretical consideration. (See Appendix \ref{app:localization}.)
%
%
For $M=2$, the classical diffusion is still suppressed, but the numerical observation tells
that localization length is enhanced and increases exponentially 
as $\ell \propto \exp \{D_{cl}N^2\}$ \cite{manai15,yamada18}.
 (See Appendix \ref{app:localization}.)

\begin{figure}[htbp]
\begin{center}
\includegraphics[width=4.0cm]{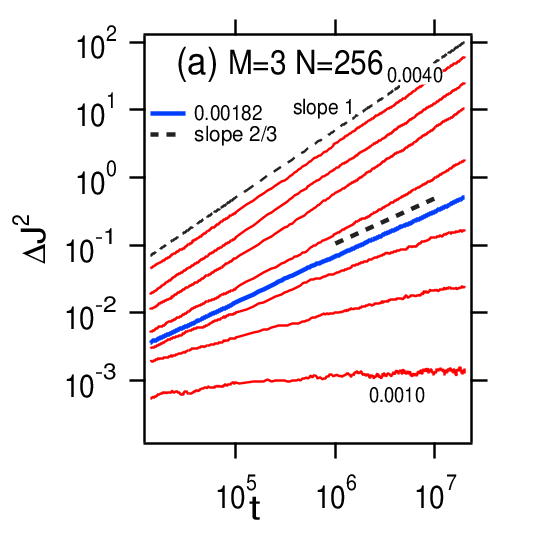}
\hspace{-2mm}
\includegraphics[width=4.0cm]{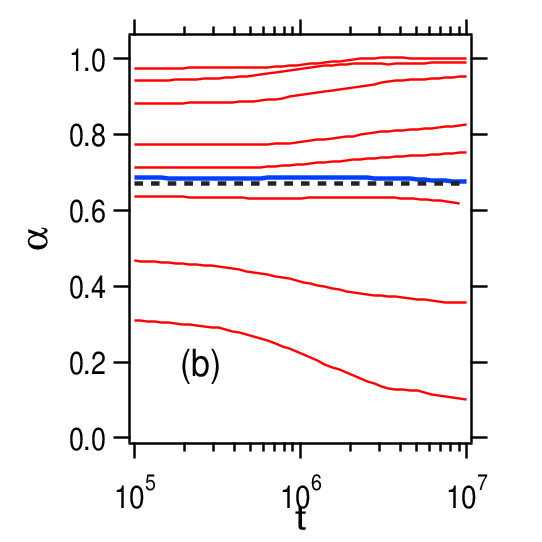}
\hspace{8mm}
\includegraphics[width=5.0cm]{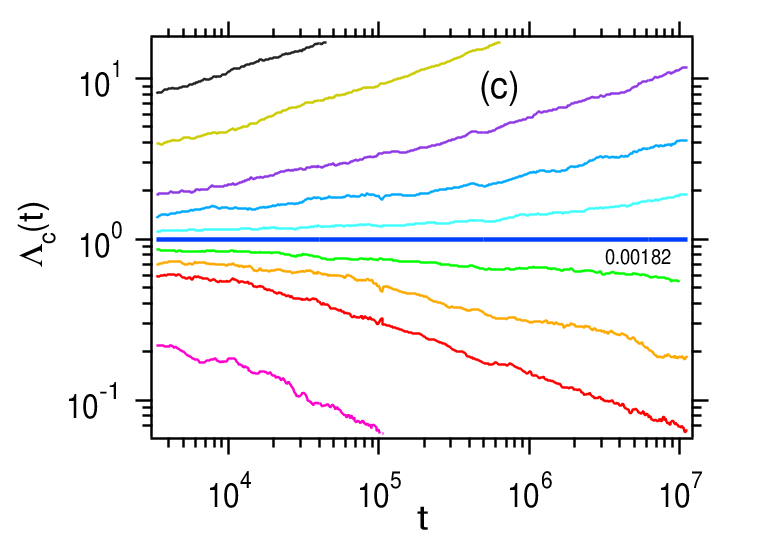}
\caption{(Color online) \label{fig:LDT}
(a)The double-logarithmic plots of $\Delta J^2(t)$ as a function of time
for the perturbed cat map for $K=-1$ of $N=256$
with trichromatic perturbation ($M=3$).
The results for $\eta_c \simeq 0.00182$ is shown in thick blue line.
The broken lines with the slope 1 and 2/3 are shown.  
(b)The instantaneous diffusion index $\alpha(t)$ for some $\eps$.
The broken line  indicates the critical subdiffusion line
$\alpha(t)=\alpha_c=2/3$ predicted by the scaling theory.
(c) The scaled MSD $\Lambda_c(t)$ 
as functions of time for increasing perturbation strengths. 
The range is $\eta \in[0.001, 0.004]$.
}
\end{center}
\end{figure}

As $M\geq 3$, things changes drastically: for small enough $\eta$ the MSD
still saturates at a finite level and the quantum localization still remains, but as
$\eta$ is taken large enough the MSD increases linearly without limit at least for 
$t$ less than $10^8$. 
Figure \ref{fig:N=8}(b) shows that such a drastic change is observed 
even for very small main system with $N$ of
only 8. The border between the localization and the 
normal diffusion is not, however, very definite.
(See Appendix \ref{app:small-N}.)
%
On the other hand, as $N$ increases greater than $10^2$
the transition from localization to the normal diffusion becomes very definite.
Figure \ref{fig:LDT} presents a typical example of $N=256$. 
There exists a critical value 
$\eta=\eta_c$ below which the MSD saturates and above which the MSD increases to 
reach to the normal diffusion. And just at $\eta=\eta_c$ the MSD increases according 
to an anomalous diffusion law $\Delta J^2(t) \propto t^\alpha$ 
with a characteristic exponent $\alpha~~(0\le \alpha\le 1)$. 
Figure \ref{fig:LDT}(b) shows the temporal behavior of MSD around $\eta=\eta_c$
by using the time-dependent characteristic exponent $\alpha(t)$ defined by
\beq
\label{eq:alpha}
  \alpha(t)=\frac{d\log \ovl{\Delta J^2(t)}}{d\log(t)}, 
\eeq
where the over-line $\ovl{X(t)}$ means to take a local time 
average of $X(t)$. Eq.(\ref{eq:alpha})
implies $\ovl{\Delta J^2(t)}$ increases $t^{\alpha(t)}$ locally at $t$. 

Figure \ref{fig:LDT}(b) shows the $(t,\alpha(t))$-plot for various $\eta$.
Below $\eta_c$, $\alpha(t)$ decreases monotonically to zero, while it increases 
to reach the normal diffusion $\alpha=1$ above $\eta_c$, which 
provides a strong evidence that a definite
transition from the localization to the normal diffusion without limit takes place.  
Just at $\eta=\eta_c$, $\alpha(t)$ takes a constant value
$\alpha_c$ and MSD exhibits an anomalous diffusion 
$\Delta J^2_{\eta=\eta_c}(t) \propto t^{\alpha_c}$. 
As $\eta$ exceeds $\eta_c$, the diffusion constant 
approaches to the classical diffusion constant $D_{cl}\propto \eta^2$. 

Once $\eta_c$ is decided by the $(t,\alpha(t))$-plot, the critical behavior close 
to $\eta_c$ can be more directly captured by the scaled representation of MSD 
$\Lambda_c(t):=\Delta J^2(t)/\Delta J^2_{\eta=\eta_c}(t)$
as shown in Fig.\ref{fig:LDT}(c). 
The critical value $\eta_c$ decreases very rapidly with $N$, which will 
be discussed later.

\begin{figure}[htbp]
\begin{center}
\includegraphics[height=4.0cm]{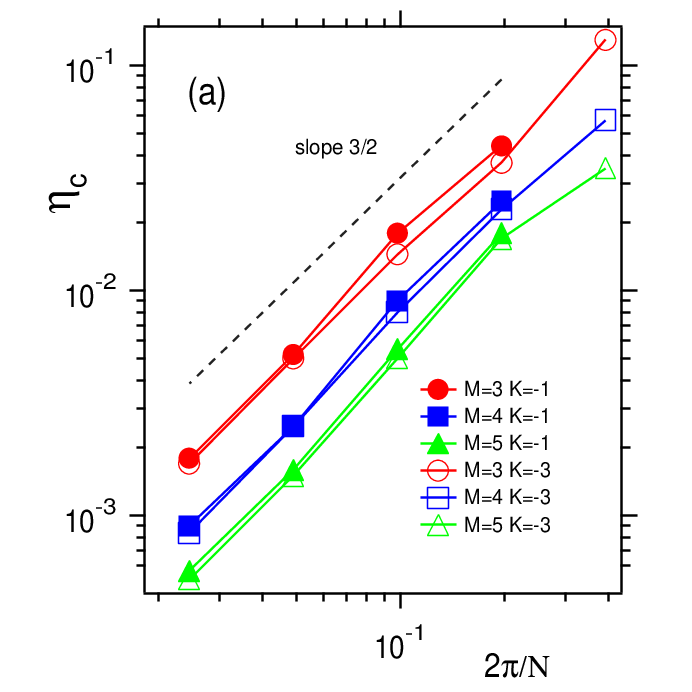}
\hspace{-4mm}
\includegraphics[height=4.0cm]{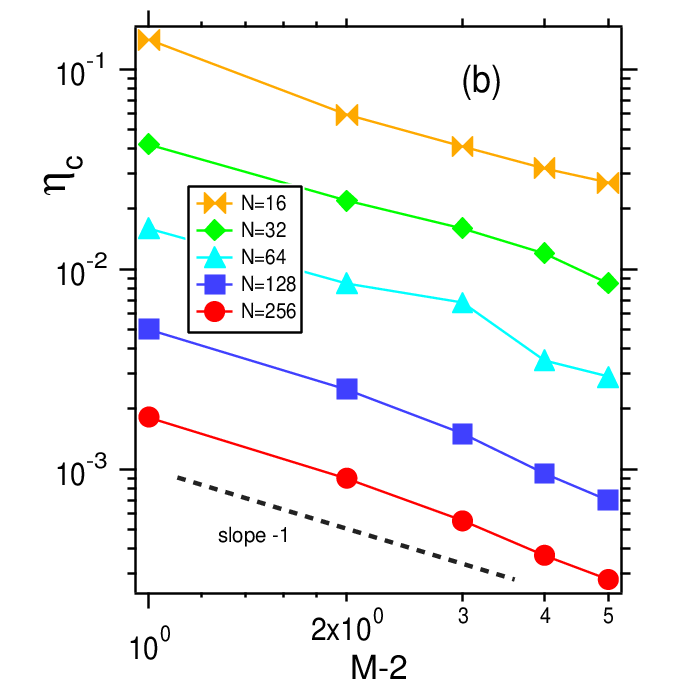}
\caption{(Color online) \label{fig:etac}
(a)The critical perturbation strength $\eta_c$  
 as a function of  $\hbar=2\pi/N$ for the perturbed cat map 
 with $M=3,4,5$ and $K=-1,-3$.
 The black broken line with a slope $3/2$ is shown.
(b)$\eta_c$ as a function of  $M-2$ for  $K=-1$.
The black dotted line with a slope $-1$ is shown.
Note that the data are plotted in double-logarithmic scale.
}
\end{center}
\end{figure}


The localization-diffusion transition is always observed if $M\geq 3$
and the MSD increases according to the subdiffusion $\Delta J^2(t) \propto t^{\alpha_c}$ at the
critical $\eta_c$, 
which decreases as $\eta_c\sim N^{-3/2}$, as shown in Fig.\ref{fig:etac}(a),
if $N\gg 1$.
 According to the numerical
observation $\alpha_c$ is independent of $N$ and depends only on
$M$ and decreases to zero with $M$. The critical value $\eta_c$ also decreases 
with $M$ as shown in Fig.\ref{fig:etac}(b). The results are summarized as  
\beq
\label{eq:critical-value}
     \alpha_c =\frac{2}{M},~~~~~~\eta_c\propto N^{-3/2}(M-2)^{-1}. 
\eeq
The diffusion phenomena can never be observed for the elliptic cat ($-4<K<1$), if $N\gg 1$
and $\eta$ is small enough.
The results mentioned above does not change if the main system is replaced by
the standard map of $|K|\gg 1$ defined on a periodically bounded phase space.

\begin{figure}[htbp]
\begin{center}
\includegraphics[height=6cm]{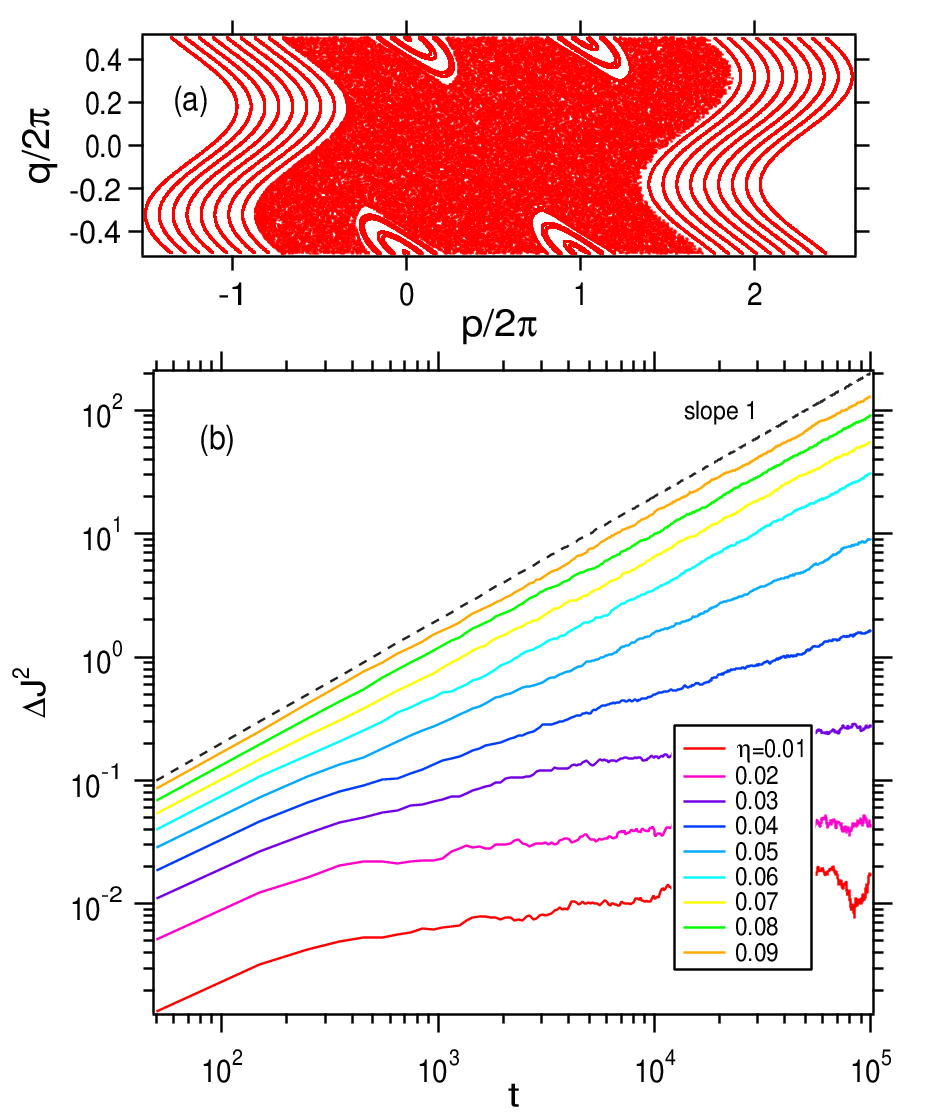}
\caption{(Color online) \label{fig:LSM}
(a) The classical Poincare map of the modified standard map,
which manifests overlapped chaos around two resonances at $p=0$ and $p=2\pi$ 
is finitely bounded by tori. Here $K=3.0$ and $M=3$, and $p_{1,2}=\pi\pm 2\pi$.
(b) Double logarithmic plots of MSD vs $t$ for (a) are shown.
Diffusion in the additional modes is recovered with increase in $\eta$,
where $N=256 (\hbar=2\pi/N)$. 
}
\end{center}
\end{figure}

The transition scheme through the critical subdiffusion $\Delta J^2(t)\sim t^{2/M}$
is very similar to the Anderson-like transition observed for standard map 
perturbed by multiply-periodic perturbations \cite{lopez12,lopez13,yamada20}.
It is, however, basically different from ours in that the diffusive motion
is supported by the chaotic degree of freedom itself. It is a large
quantum chaotic system defined in an infinitely extended phase space
and is supported by infinite dimensional Hilbert space.\\
{\it An alternative model.-} 
Next, we examine a more natural case where
a non-ideal chaos  
generated by the overlap of two nonlinear resonances is bounded finitely 
not by the periodic boundaries but
 by regular orbits. We take a modified standard map defined as follows. 
The potential is $V(\hq)=K\cos \hq$ as usual, and the momentum space is unbounded.
But the kinetic energy takes the ordinary
form $T(\hp)=\hp^2/2$ only in a bounded region $I:=[p_1,p_2]$, and $T(\hp)$ is replaced by 
linear functions $a\hp+b$ out of $I$.
The constants $a$ and $b$ are decided such that $T(\hp)$ is continuous and smooth 
at $p_1$ and $p_2$. It is easy to show the classical motion outside of $I$
is completely integrable.
By choosing $p_1$ and $p_2$
adequately, we can confine the two resonances at $p=0$ and $p=2\pi$, which
yields chaotic motion by the overlap of resonances for
$K>1$, as is illustrated in Fig.\ref{fig:LSM}(a).
 Thus our system models typical situation of a small classical 
chaos bounded by regular regions.

We examined the above system by our method. The obtained results is not so simple as
the ideal case of the cat map, but we confirmed that the normal diffusion is recovered 
for $M\geq 3$ following a similar scenario as the cat map. We show in Fig.\ref{fig:LSM}(b) the
transition process of MSD 
together with the classical Poincare plot of the main system. 

{\it Conclusion.-}
In conclusion we investigated whether small quantum chaos can induce quantum diffusion 
leading to the global transportation. As a simplest model we proposed a small but strong 
quantum chaos system coupled very weakly with additional linear modes.
The existence of diffusion depends seriously on the number $M$, and for $M\geq 3$
global diffusion is induced even for small Hilbert space dimension $N$.
IF $N\gg 1$ the diffusion is realized through a critical state exhibiting an 
anomalous diffusion as the coupling strength exceeds a weak quantum critical value. 
%
In the present work we examined sufficiently unstable chaotic systems 
in the classical limit as the main system. 
More interesting is the case of small and weakly unstable chaos typically exemplified by
the stochastic layer. In the latter case the global transportation process along the linear modes 
corresponds to a quantum Arnold diffusion, and its existence is a quite interesting problem.

We finally note that our system may be implemented as molecular vibration or rotation resonantly
excited by a pulsed laser light. If a few number of additional weak cw laser light with different 
frequencies are applied as seeds, scattering of "quantum chaotic" light to the applied 
frequency modes are induced as a quantum phase transition.  
It is a novel kind of light scattering phenomenon. 

\appendix

\section{The twinning method}
\label{app:twin-method}

\subsection{Details of the derivation of Eqs.(4)-(7)}


As presented in Eq.(2) in the main text, we decouple the action($\vc{\hJ}$)-dependent part 
from the unitary evolution operator of the Hamiltonian (1) in the main text. Indeed,
if we suppose the decoupled form $\hU=\e^{-i\hh(\vc{\hJ})t/\hbar}\hUeff$,
Schr\"{o}dinger equation becomes
\beq
\nn   \fracd{d\hUeff(t, \vc{\hphi})}{dt}&=&-\frac{i}{\hbar}
 \e^{-i\hh(\vc{\hJ})t/\hbar}\hHeff(\hp,\hq,\vc{\hphi},t) \e^{-i\hh(\vc{\hJ})t/\hbar}\hUeff(t, \vc{\hphi})\\
\nn &=& \hHeff(\hp,\hq,\vc{\hphi}+\vc{\omega}t,t)\hUeff(t, \vc{\hphi})
\eeq
and $\hUeff(t, \vc{\hphi})$ takes the form of Eq.(2)  in the main text. The equation of motion
of the action operator $\hJ_k(t)= \hU^\dagger\hJ_k\hU =\hUeff^\dagger\hJ_k\hUeff$ is given by
\beq
\nn \fracd{d\hJ_k(t)}{dt} &=& \frac{i}{\hbar}[\hHeff(\hp,\hq,\vc{\hphi}+\vc{\omega}t,t),J_k(t)] \\
\nn &=& -\eta w(\hphi_k+\omega_kt)v(\hq(t))\Delta(t).
\eeq
Thus $\hJ_k$ changes at $t=n(\in \Z)$ when the kick force is applied,
where $\hq(t)$ does not change during the kick interaction. The change of $J_k(t)$ at $t=n+0$
is easily evaluated, and summing all the changes up to $[t]$, Eq.(4) in the main text follows.\\
\\
Next we consider the dynamics of the twins. We decompose the action dependent part from
the unitary evolution operator in the same manner as follows:
\beq
\nn  \hU^{T}_\xi(t):&=& \exp\{-\frac{i}{\hbar}\int_0^t\hHeffT_\xi(s)ds \} \\
\nn    &=& \e^{-i\hh(\vc{J})t/\hbar}\hUeffT_\xi(\vc{\hphi},t),
\eeq
where 
\beq
\nn \hUeffT_\xi(\vc{\hphi},t)=\exp \{-\frac{i}{\hbar}\int_0^t \hHeffT_\xi(\hp,\hq,\vc{\phi}+\vc{\omega}s,s)ds \}.
\eeq
The transition operators $\hR^{\pm}$ , which evolves as
\beq
\hR^{\pm}(t) 
\nn &=& [\hU^{T}_\xi(t)]^\dagger\hR^{\pm}\hU^T_\xi(t) \\
\nn &=& [\hUeffT_\xi(\vc{\hphi},t)]^\dagger\hR^{\pm}\hUeffT_\xi(\vc{\hphi},t),
\eeq
commutes with the non-interacting part of the twin Hamiltonian $\hHeffT_{\xi=0}$, because
it is formed of twins and is expanded as 
\beq
\nn    \hHeffT_{\xi=0}=\sum_{q_1,q_2}\hat{A}(\vc{\hphi},q_1,q_2)[|q_1\>\<q_2|+|q_1\>'\<q_2|'].
\eeq
Thus  $\hR^{\pm}(t)$ changes by the inter-twin interaction $\hW$, and
the equation of motion leads to
\beq
\fracd{d\hR^+(t)}{dt} 
\nn &=&\xi\frac{i}{\hbar}[\hHeffT_\xi(\hp(t),\hq(t),\vc{\hphi}+\vc{\omega},t), hR^+(t)]\\
\nn &=&\xi\frac{i}{\hbar}w'(\hphi_k+\omega_kt)(v(\hq(t)-v(\hq'(t))\Delta(t-0).
\eeq
The time-dependent operator $\hX(t)$ is evolved 
in time by the action-free Hamiltonian $\hHeffT_\xi(t)$ as
$\hX(t)=[\hUeffT_\xi(t)]^{\dagger}\hX \hUeffT_\xi(t)$, 
 which contain only the angle $\vc{\hphi}$ 
operators with regard to the harmonic degrees of freedom.
Since the parameter $\xi$ is taken very small, the change of $\hR(t)$ during the
interaction process is of $O(\xi)$, and the changes of $\hq(t)$ and $\hq'(t)$ in the RHS,
 which contributes $O(\xi^2)$ correction, can be neglected. Therefore, 
\beq
 \nn & &   \hR^+(n+1)-\hR^+(n)  \\
\nn  &=&\xi\frac{i}{\hbar}w'(\hphi_k+\omega_kt)(v(\hq(n))-v(\hq'(n)),
\eeq
which yields Eq.(6) in the main text. In accordance with the approximation,
we should neglect the coupling of $O(\xi)$ in the time evolution of $\hq(n)$ and $\hq'(n)$
in the LHS of the above equation or the Eq.(6) in the text.
Then each system evolves with each Hamiltonian $\hHeff(\hp,\hq,\phi,t)$ 
and $\hHeff(\hq',\hp',\phi,t)$ respectively. 
Further the second system is unpopulated at $t=0$. Under such condition $v(\hq')$ can be neglected
and the correspondence, which is represented correctly as
\beq
\nn  
& &\fracd{\left([\hUeff(\vc{\hphi},t)]^\dagger\hJ\hUeff(\vc{\hphi},t)  - \hJ \right)}{\eta} \\
\nn~~~&=& -\fracd{i\hbar \left([\hUeffT_\xi(\vc{\hphi},t)]^{\dagger}\hR^+\hUeffT_\xi(\vc{\hphi},t) - \hR^+\right)}{\xi},
\eeq
holds between Eq.(3) and Eq.(6) in the main text. 
In the above equation the LHS is evolved by the first part of the twins alone,
but in the RHS the the time evolution is due to the weakly coupling twins.
This correspondence can be applied to compute arbitrary order moment of  $\hJ_k(t)-\hJ_k(0)$.

The initial state is taken as $|\Psi_0\>=|\vc{J}=0\>|\psi_0\>$, and the state of the cat
 $|\psi_0\>$ contains only the 
first part, and $\hR^-|\psi_0\>=\<\psi_0|\hR^+=0$. Then 
\begin{widetext}
\beq
\nn \fracd{\<\Psi_0|(\hJ(t)  - \hJ)^2|\Psi_0\>}{\eta^2} 
= \fracd{\hbar^2\<\Psi_0|([\hUeffT_\xi(t)]^{\dagger}\hR^-\hUeffT_\xi(t) - \hR^-)
([\hUeffT_\xi(t)]^{\dagger}\hR^+\hUeffT_\xi(t) - \hR^+)|\Psi_0\>}{\xi^2}.
\eeq
\end{widetext}
Further the angle-representation of the action eigenstate is
\beq
\nn  |\vc{J}=0\>=
\int_0^{2\pi}...\int_0^{2\pi}|\vc{\phi}\>d\vc{\phi}/(2\pi)^{M/2}, 
\eeq
where $d\vc{\phi}=\Pi_i^Md\phi_i$.
We thus have
\begin{widetext}
\beq
 \nn \fracd{\<\Psi_0|(\hJ(t)  - \hJ)^2|\Psi_0\>}{\eta^2} 
= \fracd{\hbar^2}{\xi^2(2\pi)^M}\int_0^{2\pi}...\int_0^{2\pi} d\vc{\phi} \<\psi_0|\<\vc{\phi}|[\hUeffT(\vc{\hphi},t)]^{\dagger}\hR^-\hR^+\hUeffT(\vc{\hphi},t) |\vc{\phi}\>|\psi_0\>.
\eeq
\end{widetext}
This results in Eq.(8) in the main text.

\begin{figure}[htbp]
\begin{center}
\includegraphics[height=5.5cm]{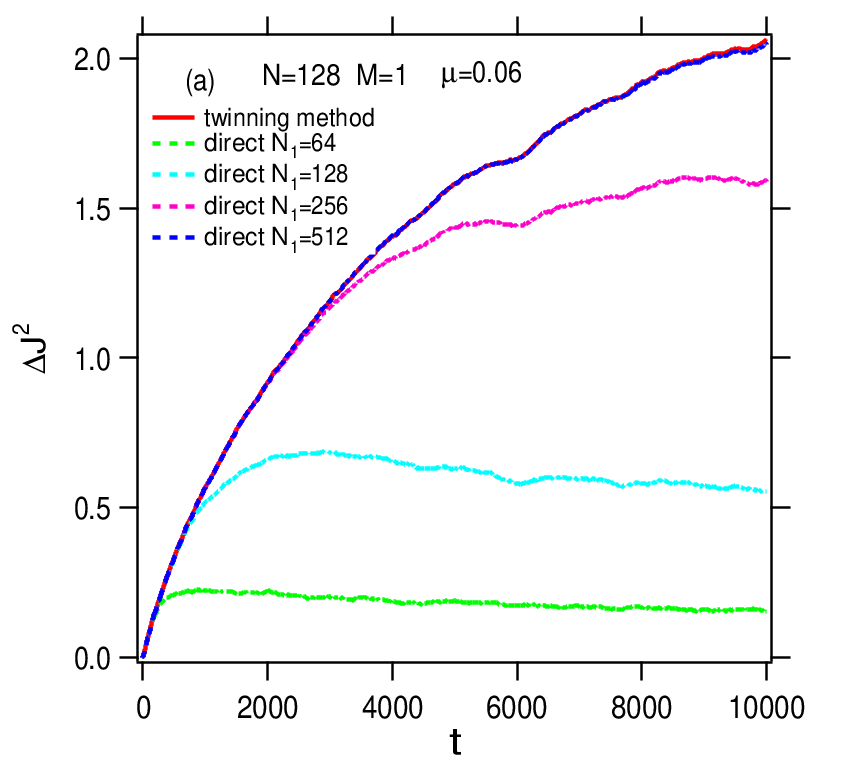}
\hspace{8mm}
\includegraphics[height=5.5cm]{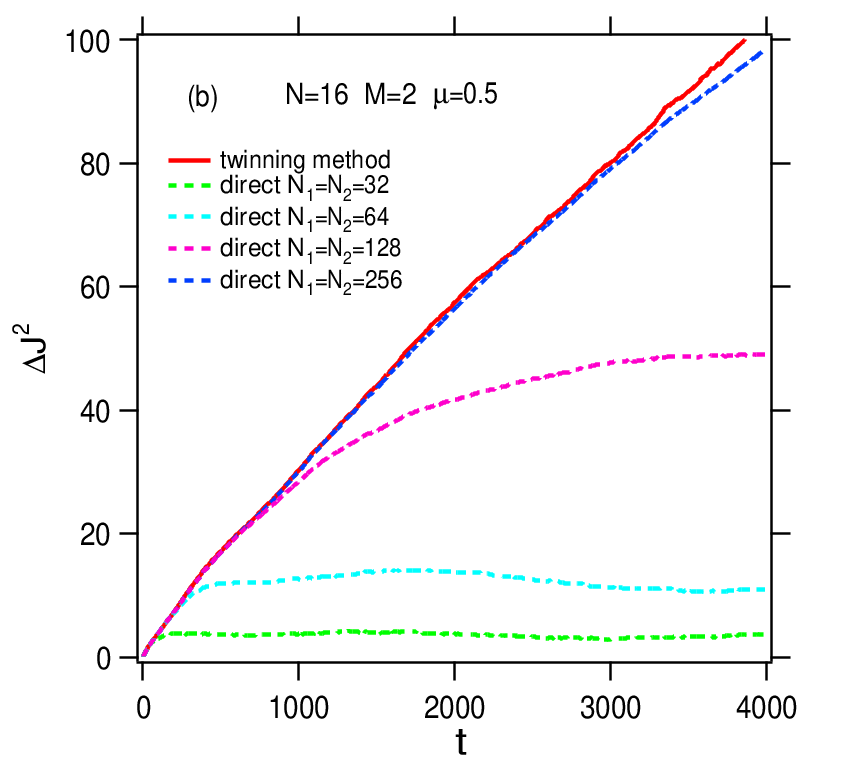}
\caption{(Color online) \label{SMFig1-method}
Comparison between the results using the twinning mehod 
and the direct method 
in the perturbed cat map  of $K=-1$.
(a)$\Delta J^2$ as a function of time for $\eta=0.06$ and $N=128$
in the monochromatically perturbed cat map ($M=1$).
$N_1=64,128,256,512$.
(b)$\Delta J^2$ as a function of time for $\eta=0.5$ and $N=16$
in the dichromatically perturbed cat map ($M=2$).
$N_1=N_2=32,64,128,256$.
}
\end{center}
\end{figure}

\begin{figure}[htbp]
\begin{center}
\includegraphics[height=5.5cm]{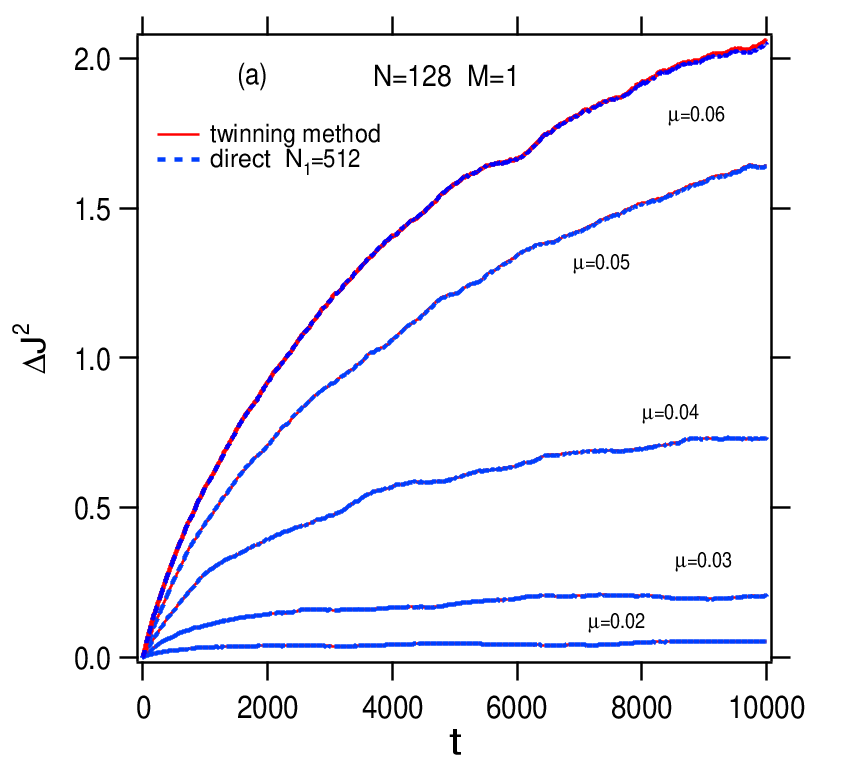}
\hspace{8mm}
\includegraphics[height=5.5cm]{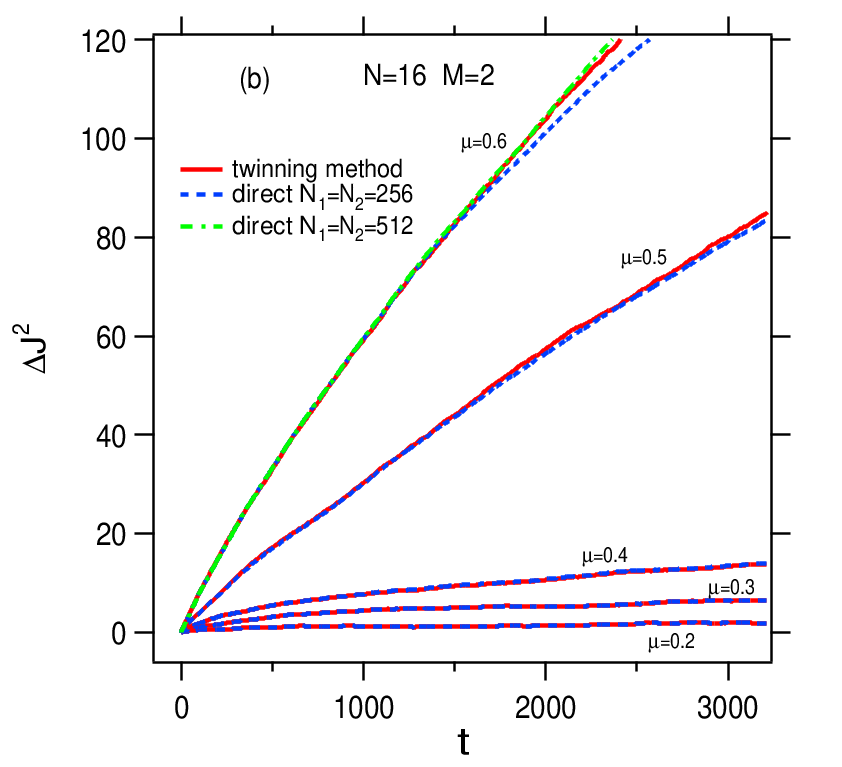}
\caption{(Color online) \label{SMFig2-result}
$\Delta J^2$ as a function of time for various $\eta$ 
in (a)the monochromatically perturbed cat map ($M=1$) of $K=-1$ and $N=2^7$
and (b)in the dichromatically perturbed cat map ($M=2$) of $K=-1$ and $N=2^4$. 
The numerical results by using the twinning method and the direct method using FFT
 are shown by red solid line and blue dashed lines, respectively.
Note that the all axes are in the linear scale.
}
\end{center}
\end{figure}

\subsection{Numerical check of the twining method}
We shall numerically compare the results obtained by the twinning method 
with those  obtained by the usual method of the wavepacket  propagation
in the action space of the linear subsystem, where the $M-$dimensional action
space is spanned by the set truncated action basis with the Hilbert-space 
dimension $N_i~(1\leq i\leq M)$.

Figure \ref{SMFig1-method} shows the $\Delta J^2$ as a function of time for
$M=1$ and $M=2$ with a given set of parameters.
It can be seen that as the number of Hilbert-space dimension
$N_1$ and $N_2$ of the action spaces
increases,  the results approaches to those obtained by the twin method 
,where the FFT-method is applied in the direct wavepacket propagation.

Figure \ref{SMFig2-result} shows the $\Delta J^2$ as a function of time for
various coupling strength $\eta$ in the cases of $M=1$ and $M=2$ for
large enough $N_1=N_2$. It can be seen that the results from both methods 
agree well.  

If diffusion like behavior takes place in the linear subsystem,  the truncated dimension 
$N_1, N_2,...N_M$ must be increased in proportion to the diffusion length $\propto t^{1/2}$.
as the time-scale of simulation $t$ increases.  And the direct method requires the 
CPU time $\propto t^{1+M/2}$. On the other hand, the CPU time required for the
twin method increases only as $\propto t$.
Incidentally the CPU time required for the twin method is $70$ seconds, while
the direct method requires  $36\times 10^3$ seconds for $N_1=N_2=256$  for the results of 
the Fig.\ref{SMFig1-method}(b).
These facts truly demonstrate the usefulness of the twin method. However, we also note 
that the twin method is applicable only for the case of  linear subsystems.
That is, we can call it poor man's method for poor man's model.

\section{Localization length for $M=1$ and $M=2$}
\label{app:localization}
%
%
%

\begin{figure}[htbp]
\begin{center}
\includegraphics[height=5.5cm]{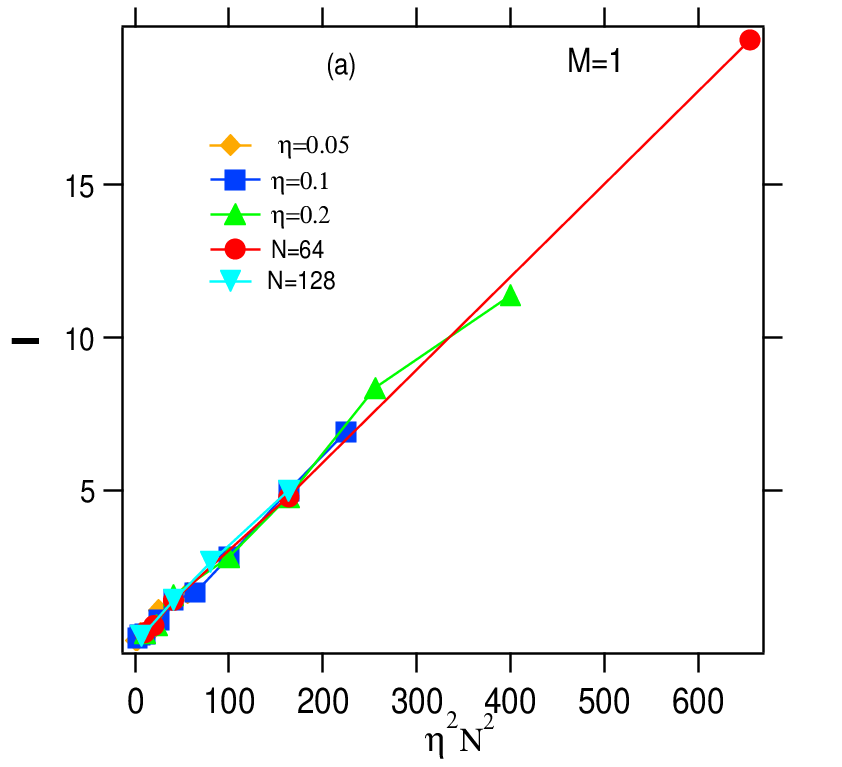}
\hspace{8mm}
\includegraphics[height=5.5cm]{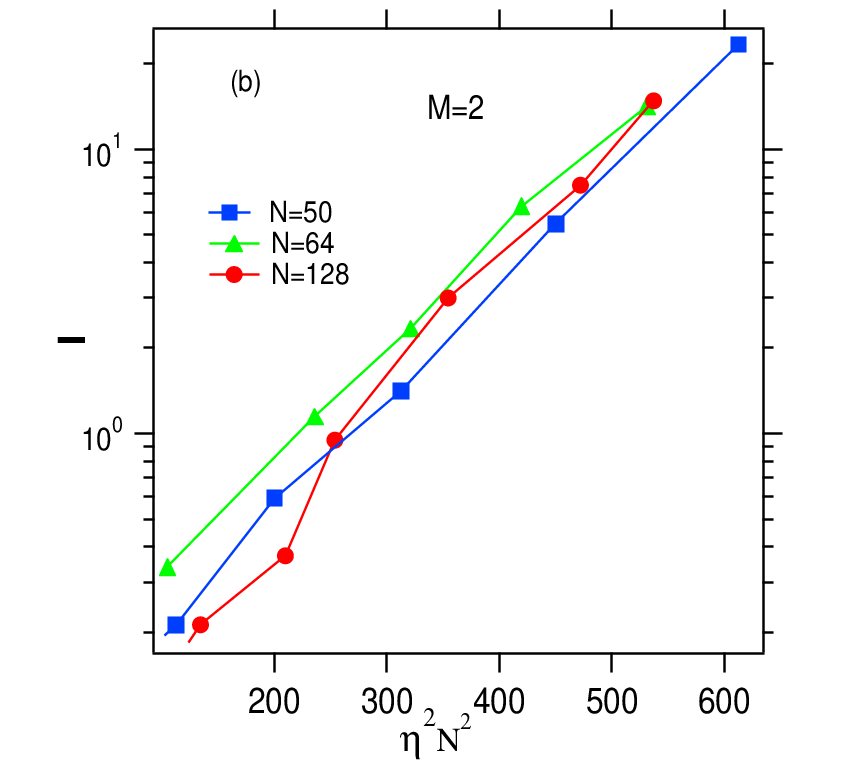}
\caption{(Color online) \label{SMFig3-loc}
Dynamical localization length of the linear mode $J_1$
as a function of $\eta^2N^2$($\propto \eta^2D_{cl}$) for some parameters 
in the perturbed cases $M=1$ (a) and $M=2$ (b).  
Note that the vertical axis is in the linear scale for $M=1$ 
and in the logarithmic scale for $M=2$, respectively.
}
\end{center}
\end{figure}

\def\Dcl{D_{cl}}
Some properties of localization length $\ell$ characterizing the dynamical localization
observed for $M=1$ and $2$ are presented when the main system is hyperbolic Arnold cat map.  
Prior to showing the results, we briefly comment on the classical dynamics of the linear systems.
Regarding Eq.(4) as the classical equation, the $n$-step local diffusion constant
$\Dcl(n)$ can be defined by
$$
 \nn  \Dcl(n)  =\eta^2[c(n,n)+2\sum_{n'<n} c(n,n')],
$$
and 
$$
 \nn (J(t)-J(0))^2 = \sum_{n\leq [t]}  \Dcl(n) \\
$$
where $c(n,n')= w_k'(\omega_k n)w_k'(\omega_k n')v(q_n)v(q_{n'})$ is correlation function. 
If the main system is fully chaotic and $\eta\ll 1$, the feature of chaos of the main system
is not modified by the perturbation of linear modes, and  for $t\to \infty$
the average $\Dcl:=\sum_{n=0}^{[t]}\Dcl(n)/[t]$ 
converges to give the classical diffusion constant which obeys $\propto \eta^2$. 
In particular for the hyperbolic 
cat map the correlation function decays suddenly to give 
$$
\Dcl=\eta^2/4
$$
if $w_k(\phi)=\cos(\phi)$ and $v(q)=cos(q)$.  
These facts can easily be confirmed by numerical simulation.

With this result we can conjecture that if the diffusion along a linear mode stops due to the quantum effect
the localization length $\ell$ will be proportional to the  scaled parameter $\eta^2N^2$. Let us examine the
case of $M=1$. If the diffusion stops at $n_s$-step with the localization length $\ell$, the classical
results leads to $n_s\Dcl=\ell^2$. The number of quantum state related to the suppressed diffusion is 
$N_{loc}=N\times[$ number of action states in $\ell]=N\ell/\hbar$.
 Since the interval of quantum eigen angle is roughly $2\pi/N_{loc}$, 
 the diffusion terminates at $n_s \sim N_{loc}$. 
These relations results into
$$
  \ell \propto N^2\eta^2 \sim \Dcl N^2~~~{\rm for }~~~M=1.
$$
The numerical results shown in Figure \ref{SMFig3-loc}(a) agrees quite with the predicted dependency
on the combined parameter $\eta^2N^2 \sim \Dcl N^2$.

The above intuitive argument can not be extended to the case of $M=2$. The numerical results
reveals a quite different  nature of $\ell$ which increases exponentially with $\Dcl$ and $N$.
However, the combined parameter $\eta^2N^2$ still plays a central role for $M=2$.
Figure \ref{SMFig3-loc}(b) shows that  exponential growth of the localization length 
is also almost controlled by the combined  parameter $N^2\Dcl$.
As a result, the dynamical localization length can be summarized as follows: 
$$
 \ell \propto \e^{c_1\eta^2 \Dcl} ~~~{\rm for }~~~M=2, 
$$
where  $c_1$ is a numerical constant.
Similar phenomena have been observed in large quantum chaos system with infinite Hilbert-dimension
such as perturbed standard map \cite{manai15} and  perturbed Anderson map \cite{yamada18}.

\section{Transition to diffusion in case of small  $N$ and $M=3$}
\label{app:small-N}
\begin{figure}[htbp]
\begin{center}
\includegraphics[height=5.5cm]{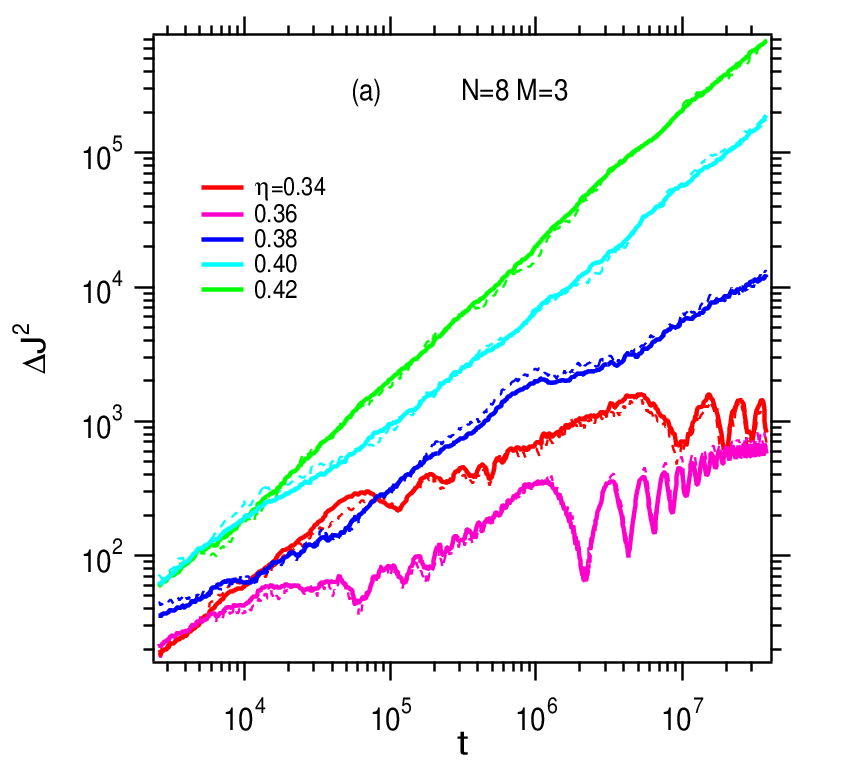} 
\hspace{8mm}
\includegraphics[height=5.5cm]{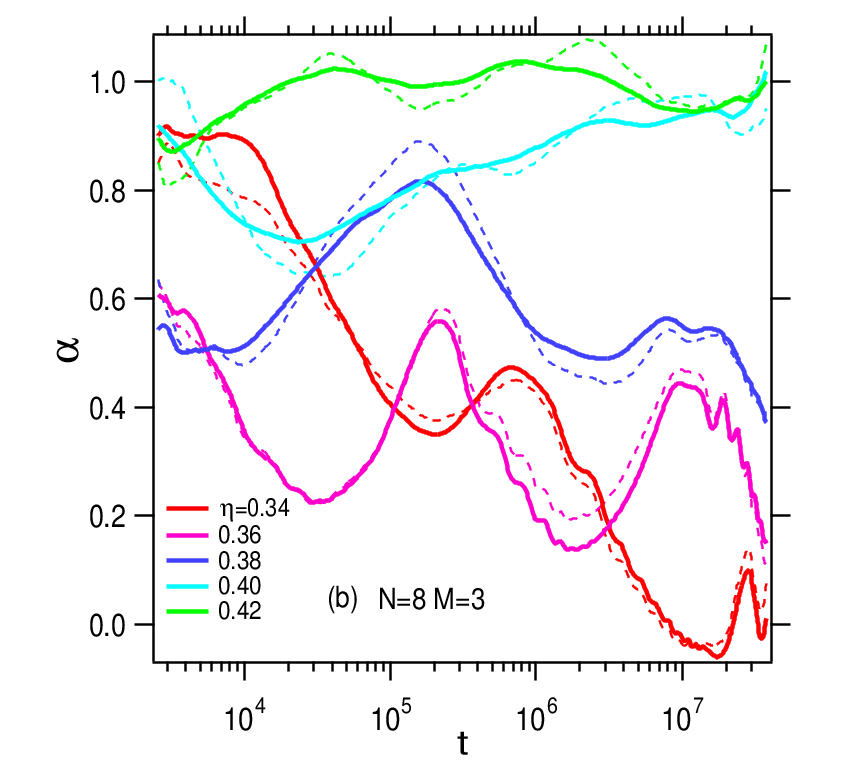}
\caption{(Color online) \label{SMFig4-result} 
(a)$\Delta J^2$ as a function of time for various $\eta$
in the trichromatically perturbed cat map ($M=3$) of $K=-1$ and $N=2^3$.
(b)Plots of $\alpha$ as the function of $t$ for various $\eta$.
The results for a single initial state $|\phi_0\>$ (by broken lines)
and the average over 8 orthogonal random $|\phi_0\>$ (by thick solid lines)
are compared.
The patterns of fluctuation almost coincide.
}
\end{center}
\end{figure}

As shown in Fig.1(b) in the main text, a transition 
to diffusion takes place with increase of $\eta$ even if
the Hilbert space dimension $N$ is not large.  The nature of the transition is, however, 
different from the case of $N\gg1$ in which a clear critical state with a definite critical 
index $\alpha_c<1$ exists at a critical value $\eta=\eta_c$.
Instead, there seems to be a transition range of $\eta$:  
if $\eta$ is taken in the transition range, the time-dependent index $\alpha(t)$ 
defined by Eq.(8) fluctuates notably as a
function of time around a certain value 
$\ovl{\alpha}$ as is depicted in Fig.\ref{SMFig4-result} and
the mean value $\ovl{\alpha}$ tends to increase with $\eta$ and 
 approaches the line $\alpha=1$ beyond 0.4.  
For the different choice of initial state, the pattern of the fluctuation is quite similar 
as shown in Fig.\ref{SMFig4-result}.
It means that the  insensitivity of the fluctuation to the initial state 
and has a quantum origin. 

The occurrence of stationary diffusion implies the correlation function $c(n,n')$ 
in the last section converges and loss of memory is maintained.
The mechanism by which such an apparently irreversible behavior is self-organized  in 
integrable systems by coupling with very small quantum chaos consisting of 
only a few quantum states  is an important issue. We have currently no theoretical tools
to attack the issue.

\begin{acknowledgements}
This work is supported by Japanese people's tax via 
MEXT/JSPS KAKENHI Grant Number 22K03476 and 22H01146,
and the authors would like to acknowledge them.
The authors are also thankful to Kankikai (Dr. T.Tsuji) 
and  Koike memorial house for using the facilities 
during this study.
\end{acknowledgements}



\begin{thebibliography}{00}

\bibitem{lichtenberg92}
A.J. Lichtenberg and M.A. Lieberman,
{\it Regular and Chaotic Dynamics} (Applied Mathematical Sciences 38),
(Springer; 2nd ed. 1992).

\bibitem{chirikov79}
B.V.Chirikov, 
 Phys. Rep. {\bf 52} 263 (1979).
 
\bibitem{arnold64}
V.I. Arnol'd,
Soviet Mathematics-Doklady 
{\bf 5}, 581-5(1964).



\bibitem{nekhoroshev77}
N. N. Nekhoroshev, 
Usp. Mat. Nauk {\bf 32}, 5 (1977).

\bibitem{lieberman80}
M. A. Lieberman, 
Ann. N. Y. Acad. Sci. {\bf 357}, 119 (1980). 

\bibitem{holmes82}
P. J. Holmes and J. E. Marsden,
J. Math. Phys. {\bf 23}, 669(1982).



\bibitem{kaneko85}
K. Kaneko and Richard J. Bagley,
Phys.Lett. A {\bf 110} 435-440(1985).






\bibitem{cincotta14}
P.M.Cincotta {\it et al.}, 
Physica D {\bf 266}, 49-64(2014). 

\bibitem{santhanam22}
M.S. Santhanam, S. Paul, and J. BharathiKannan,
Physics Reports {\bf 956}, 1-87(2022).

\bibitem{maurya22}
S. S. Maurya, 
J. B. Kannan, K. Patel, 
P. Dutta, K. Biswas, J. Mangaonkar, M. S. Santhanam, and U.D. Rapol,
Phys. Rev. E {\bf 106}, 034207(2022).

\bibitem{boretz16}
Y.Boretz, and L. E. Reichl. 
Phys. Rev. E {\bf 93},032214(2016).










\bibitem{wolynes97}
David M. Leitner and Peter G. Wolynes,
Phys.Rev. Lett. {\bf 79}, 55-58(1997).

\bibitem{demikhovskii02a}
V.Ya. Demikhovskii, F.M. Izrailev, and A.I. Malyshev,
Phys.Rev. Lett. {\bf 88}, 154101(2002).

\bibitem{demikhovskii02b}
V.Ya. Demikhovskii, F.M. Izrailev, and A.I. Malyshev,
Phys. Rev. E {\bf 66}, 036211(2002).


\bibitem{demikhovskii06}
V.Ya. Demikhovskii, F.M. Izrailev, and A.I. Malyshev,
Phys. Lett. A {\bf 352}, 491 (2006).



\bibitem{casati79} 
G. Casati, B. V. Chirikov,F. M. Izraelev, J.Ford,
{\it Stochastic behavior of a quantum pendulum under a periodic perturbation}
(Springer-Verlag,Berlin,1979)  ed. by G.Casati and J.Ford, pp334.


\bibitem{fishman82}
S. Fishman, D.R.Grempel, R.E.Prange, 
Phys. Rev. Lett. {\bf 49}, 509 (1982).




\bibitem{casati89}
G.Casati, I.Guarneri and D.L.Shepelyansky, Phys. Rev.Lett. {\bf 62}, 345(1989).

\bibitem{chotorlishvili19}
L. Chotorlishvili, S. Stagraczynski, M. Schlerc and J. Berakdar,
ACTA PHYSICA POLONICA A {\bf 135} 1155-1161(2019).



\bibitem{adachi88b}
S. Adachi, M. Toda, and K. Ikeda,
Phys. Rev. Lett. {\bf 61}, 659(1988).


\bibitem{gadway13}
B.Gadway, J.Reeves, L.Krinner, D.Schneble
Phys. Rev. Lett. {\bf 110}, 190401(2013).










\bibitem{lopez12}
M.Lopez, J.F.Clement, P.Szriftgiser, J.C.Garreau, and D.Delande, 
Phys. Rev. Lett. {\bf 108}, 095701(2012).

\bibitem{lopez13}
M. Lopez, J.-F. Clement, G. Lemarie, D. Delande,
P. Szriftgiser, and J. C. Garreau,
New J. Phys. {\bf 15},  065013(2013).



\bibitem{yamada20}
H.S.Yamada and K.S.Ikeda
Phys. Rev. {\bf E101}, 032210 (2020).


\bibitem{shepelyansky83}
D.L. Shepelyansky, 
Physica D {\bf 8},208-222(1983).


\bibitem{chabe08}
J.Chabe,G.Lemarie,B.Gremaud, and D.Delande, P.Szrifigiser, and J.C.Garreau,
Phys. Rev. Lett. {\bf 101}, 255702(2008).

\bibitem{lemarie09}
G.Lemarie, J.Chabe, P.Szrifigiser, J.C.Garreau, B.Gremaud, and D.Delande, 
Phys. Rev. A {\bf 80}, 043626(2015).


\bibitem{manai15}
I.Manai, J.F.Clement, R.Chicireanu, C.Hainaut, J.C.Garreau,
P.Szriftgiser,  and D.Delande, 
Phys. Rev. Lett. {\bf 115}, 240603(2015).

\bibitem{yamada18}
H.S.Yamada F.Matsui, and K.S.Ikeda,
Phys. Rev. {\bf E 97}, 012210 (2018).



\end{thebibliography}
\end{document}